\documentclass[superscriptaddress,pra,twocolumn]{revtex4}

\usepackage[latin1]{inputenc}
\usepackage{graphicx}
\usepackage{amsmath}
\usepackage{amssymb}
\usepackage{dsfont}
\usepackage{graphicx}

\makeatletter
\def\erf{\mathop{\operator@font erf}\nolimits}
\makeatother

\newcommand\be{\begin{equation}}
\newcommand\ee{\end{equation}}

\begin{document}

\title{New method to evaluate divergent series via the Wigner function}

%\author{}
%\affiliation{INAOE, Apdo. Postal 51 y 216, 72000, Puebla, Pue.,
%Mexico}
\author{H\'ector Moya-Cessa,$^1$ Roberto de Jes\'us Le\'on-Montiel,$^2$ and Erwin A. Mart\'{\i}-Paname\~no$^{2,}$}
\affiliation{INAOE, Apdo. Postal 51 y 216, 72000, Puebla, Pue.,
Mexico\\$^2$Benem\'erita Univ. Aut\'onoma Puebla, Fac Ciencias
F\'{\i}s. Mat., Apartado Postal 1152, Puebla, Pue. 72000, Mexico}
\begin{abstract}
It is shown how a physical function, namely the Wigner function,
that in principle may be   measured, can be used to evaluate
divergent series.
\end{abstract}
\pacs{02.30.Lt, 02.10.De, 02.10.Ox}
\maketitle We give a physical
way to evaluate divergent series. Abel quoted on them:

"Divergent series are on the whole devil's work, and it is a shame
that one dares to found any proof on them. One can get out of them
what one wants if one uses them, and it is they which have made so
much unhappiness and so many paradoxes" \cite{Truc}.

We can evaluate the alternating series $S=1-1+1-1+1+ ...$ by
considering the alternating geometric series $1-x+ x^2 -x^3+x^4 -
...$ provided $|x| < 1$,  we know that it converges to $1/(1+x)$.
By allowing $x$ to get close to $1$, the series will tend to a
value close to $0.5$. Therefore, we can say that the (divergent)
sum $S$ evaluates $1/2$ in the Abel sense \cite{Truc}.

Another way of defining convergence is by using C\'esaro's sums.
The sum $S$ diverges not because the partial sums grow
uncontrollably, but rather because the partial sums oscillate. If
we could find a way of averaging the sums in order to smooth them
out, maybe this series will converge. Using Abel and C\'esaro's
limits it has been shown also that $1-2+3-4+5 + ...$ has discrete
sum $-1/4$ \cite{Stone}, values that agree with evaluations of the
Riemann zeta function \cite{Stone}.

In this contribution we propose a new method to evaluate divergent
series by using a physical function, namely the Wigner function
\cite{Wigner}, which may be written in the form,    \cite{Cahill}
\begin{equation}
W(\alpha)= \frac{1}{\pi}Tr\{(-1)^{\hat{n}}D^{\dagger}(\alpha )\rho
D(\alpha )\}\end{equation} where $\rho$ is the system's density
matrix, the number operator $\hat{n}=a^{\dagger}a$ with $a$ the
annihilation operator and $D(\alpha )=\exp(\alpha
a^{\dagger}-\alpha^*a)$ the Glauber displacement operator, where
$\alpha=(q+ip)/\sqrt{2}$. Another common form for the Wigner
function is
\begin{equation}
W_A(\alpha)=\frac{1}{2\pi}\int du e^{iup}\langle
q+u/2|A|q-u/2\rangle
\end{equation}
where we define it for an arbitrary operator $A$. If we consider a
function of the position operator, we find
\begin{equation}
W_{f(\hat{q})}(\alpha)=\frac{1}{2\pi}\int du
e^{iup}f(q-u/2)\delta(u)=\frac{f(q)}{2\pi}.
\end{equation}

It is worth noting that the Wigner function is in general a
physical function and that has been measured in experiments, in
particular for the first excited state of the vibrational motion
of an ion \cite{Leib}.

For an arbitrary operator, for instance, position to an arbitrary
power $\hat{q}^k$ the Wigner function may also be given in a
series representation \cite{Moyarep}
\begin{eqnarray}
\nonumber \frac{q^k}{2\pi}&=&
\frac{1}{\pi}\sum_{n=0}^{\infty}(-1)^n\langle n|
D^{\dagger}(\alpha )\hat{q}^k D(\alpha )|n\rangle \\ &=&
\frac{1}{\pi}\sum_{n=0}^{\infty}(-1)^n\langle n| (\hat{q}+q)^k
|n\rangle \label{alter}.
\end{eqnarray}
where $|n\rangle$ is a number state. Note that equation
(\ref{alter}) has already the alternating form of the series
considered above.

By doing $k=0$, we obtain
\begin{eqnarray}
\frac{1}{2}=\sum_{n=0}^{\infty}(-1)^n, \label{onehalf}
\end{eqnarray}
doing $k=2$ gives \cite{explanation}
\begin{eqnarray}
\frac{q^2}{2}=\sum_{n=0}^{\infty}(-1)^n(q^2+\langle n|
\hat{q}^2|n\rangle), \label{inter}
\end{eqnarray}
which by using that $\langle n| \hat{q}^2|n\rangle=n+1/2$ allows
us to evaluate the sum
\begin{eqnarray}
\sum_{n=0}^{\infty}(-1)^n n=-\frac{1}{4}, \label{aquarter}
\end{eqnarray}
where we have used (\ref{onehalf}) to evaluate (\ref{aquarter}).
Eq. (\ref{inter}) then shows  that there is a recursion relation
between the higher order sums in terms of the lower order ones. We
can find a general expression that will have this recursion in it.
In order to do this we write $\hat{q}$ in terms of annihilation
and creation operators,
\begin{eqnarray}
\nonumber\hat{q}=\frac{a+a^{\dagger}}{\sqrt{2}}, \qquad &&
a|n\rangle=\sqrt{n}|n-1\rangle,\\
&& a^{\dagger}|n\rangle=\sqrt{n+1}|n+1\rangle,
\end{eqnarray}
and insert it in (\ref{alter}), this gives
\begin{equation}
 \frac{q^k}{2}=
\sum_{n=0}^{\infty}(-1)^n\sum_{s=0}^k \left(\begin{array}{c} k \\
s
\end{array}\right)\frac{q^{k-s}}{2^{s/2}} \langle n| (a+a^{\dagger})^s
|n\rangle .
\end{equation}
by equating the coefficients of powers of $q$ at right and left of
the equal sign, we finally obtain ($s>0$)
\begin{eqnarray}
0=\sum_{n=0}^{\infty}(-1)^n\langle n| (a+a^{\dagger})^{2s}
|n\rangle . \label{main}
\end{eqnarray}
This equation is the main result of the manuscript, as it will
allow the evaluation of the divergent series considered earlier.

Because of the average with the numbers states, we can remove all
terms that do not contain an equal number of ${a}^\dagger$'s and
${a}$'s, as they are the only terms that will  contribute to the
sum of diagonal matrix elements. Therefore by considering those
elements we may  neglect all other terms of the sum. We have
\cite{Larson}
\begin{equation}\label{rwa1}
\left({a}^\dagger+{a}\right)^m\Rightarrow\left\{\begin{array}{lll}
\left(\begin{array}{c}m \\ m/2\end{array}\right):({a}^\dagger
{a})^{m/2}:_W, & & m\,\,\mathrm{even}\\ \\ 0\, & &
m\,\,\mathrm{odd}\end{array}\right.
\end{equation}
where $:({a}^\dagger {a})^s:_W$ denotes the Weyl (symmetric)
ordering of the operator $\hat{n}^s$. It may be transformed into
normal ordering using \cite{Fuji}
\begin{equation}\label{rwa2}
:({a}^\dagger
{a})^m:_W=\sum_{l=0}^m\frac{l!}{2^l}\left(\begin{array}{c}m \\
l\end{array}\right)^2{a}^{\dagger m-l}{a}^{m-l}.
\end{equation}
Inserting (\ref{rwa2}) into (\ref{main}) and making use of the
expression
\begin{equation}
a^{\dagger m}a^m|n\rangle=\frac{n!}{(n-m)!}|n\rangle,
\end{equation}
we obtain
\begin{equation}
\sum_{n=0}^{\infty}\sum_{l=0}^s\frac{(-1)^n}{2^l}\left(\begin{array}{c}s \\
l\end{array}\right)\left(\begin{array}{c}n \\
s-l\end{array}\right)=0.
\end{equation}
For $s=1$ we obtain (\ref{onehalf}) and for $s=2$ we obtain
(\ref{aquarter}), while for $s>2$ we obtain the sums
\begin{equation}
\sum_{n=0}^{\infty}(-1)^nn^{(s-1)}
\end{equation}
in terms of the {\it lower} sums, i.e. as a recursion formula.

In conclusion, we have shown a physical form, to evaluate some
divergent sums, by means of a function that in principle may be
measured, namely, the Wigner function.


\begin{thebibliography}{88}
\bibitem{explanation} We do not consider odd powers because
their results  are contained in the even powers evaluations.
\end{thebibliography}
\end{document}